\documentclass[amssymb,amsmath,aps,groupedaddress,superscriptaddress]{revtex4}
\usepackage{epsfig}
\usepackage{graphicx}

\newcommand{\forget}[1]{}
\def\lambdabar{\protect\@lambdabar}
\def\@lambdabar{%
\relax
\bgroup
\def\@tempa{\hbox{\raise.73\ht0
\hbox to0pt{\kern.25\wd0\vrule width.5\wd0
height.1pt depth.1pt\hss}\box0}}%
\mathchoice{\setbox0\hbox{$\displaystyle\lambda$}\@tempa}%
{\setbox0\hbox{$\textstyle\lambda$}\@tempa}%
{\setbox0\hbox{$\scriptstyle\lambda$}\@tempa}%
{\setbox0\hbox{$\scriptscriptstyle\lambda$}\@tempa}%
\egroup
}

\forget{
\def\subequations{\refstepcounter{equation}%
\edef\@savedequation{\the\c@equation}%
\edef\@savedtheequation{\the\@stequation}
\edef\oldtheequation{\theequation}%
\setcounter{equation}{0}%
\def\theequation{\oldtheequation\alph{equation}}}%

\def\endsubequations{%
\setcounter{equation}{\@savedequation}%
\edef \theequation{\the\@steq} \global }
}

\begin{document}

\title{{Quantum electrodynamics of accelerated atoms in free space and in confined cavities} }

\author{Alexey Belyanin}
\affiliation{Institute for Quantum Studies and Department of Physics, Texas A\&M Univ.,
TX 77843, USA}
\author{Federico Capasso}
 \affiliation{Division of Engineering and Applied Sciences, Harvard University,
Cambridge, Massachusetts, USA}
\author{Edward Fry}
\affiliation{Institute for Quantum Studies and Department of Physics, Texas A\&M Univ.,
TX 77843, USA}
\author{Stephen Fulling}
\affiliation{Department of Mathematics, Texas A\&M Univ., TX 77843, USA}
\author{Vitaly V. Kocharovsky}
\affiliation{Institute for Quantum Studies and Department of Physics, Texas A\&M Univ.,
TX 77843, USA} \affiliation{Institute of Applied Physics RAS, 603950 Nizhny Novgorod,
Russia}
\author{M. Suhail Zubairy}
\affiliation{Institute for Quantum Studies and Department of Physics, Texas A\&M Univ.,
TX 77843, USA}
\author{Marlan O. Scully}
\affiliation{Institute for Quantum Studies and Department of Physics, Texas A\&M Univ.,
TX 77843, USA}  \affiliation{Center for Ultrafast Laser Applications and Department of
Chemistry, Princeton University, Princeton NJ, USA}

\begin{abstract}
We consider a gedanken experiment with a beam of atoms in their ground state that are
accelerated through a single-mode microwave cavity. We show that taking into account of
the ''counter-rotating'' terms in the interaction Hamiltonian leads to the excitation
of an atom with simultaneous emission of a photon into a field mode. In the case of a
 slow switching on of the interaction, the ratio of emission and
absorption probabilities is exponentially small and is described by the Unruh factor.
In the opposite case of sharp cavity boundaries the above ratio is much greater and
radiation is produced with an intensity which can exceed the intensity of Unruh
acceleration radiation in free space by many orders of magnitude. In both cases real
photons are produced, contrary to the opinion that a uniformly accelerated atom does
not radiate. The cavity field at steady state is described by a thermal density matrix.
However, under some conditions laser gain is possible. We present a detailed discussion
of how the acceleration of atoms affects the generated cavity field in different
situations, progressing from a simple physical picture of Unruh radiation to more
complicated situations.
\end{abstract}
\maketitle

\section{Introduction}

Intriguing properties of vacuum as viewed by accelerated observers have been the
subject of intense investigation for almost three decades. One of the most remarkable
results is the so-called Fulling-Unruh effect predicted and analyzed by Davies,
Fulling, Unruh and DeWitt \cite{Davies}, and others \cite{2-7}. In essence, it was
shown that ground state atoms, accelerated through vacuum, are promoted to an excited
state just as if they were in contact with a blackbody thermal field. These studies
predict that a (two-level) ground state atom, having transition frequency $\omega$, and
experiencing a constant acceleration $a$, will be excited to its upper level with a
probability governed by the Boltzmann factor ${\rm{exp}} (-2\pi \omega /\alpha)$, where
$\alpha=a/c$, $c$ is the speed of light in vacuum. Unfortunately, even for very large
acceleration ``frequency" $\alpha$ $\approx 10^{8}$ Hz \cite{8}, and microwave
frequency $\omega$ $\approx10^{10}$ Hz \cite{9}, this factor is exponentially small,
$\sim 10^{-200}$; and is not of experimental interest.

To begin with, we note that the physical picture of the Unruh effect in any setting is
quite straightforward. In particular, it is the counter-rotating terms in the
interaction Hamiltonian that describe the process of an excitation of an atom with
simultaneous emission of a photon (see Fig.~1).

\begin{figure}[ht]
\centerline{\includegraphics[scale=0.6]{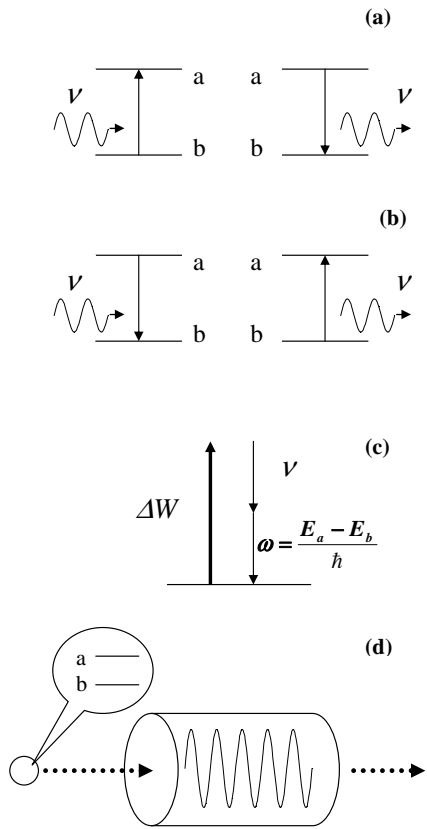}} \caption{\label{Fig01}  (a) Resonant
absorption or emission: an atom is excited (deexcited) as it simultaneously absorbs
(emits) a photon. (b) Counter-resonant absorption or emission processes that are
usually neglected in the ``rotating wave approximation'': an atom is excited
(deexcited) as it simultaneously emits (absorbs) a photon. (c) The energy for
counter-resonant emission is drawn from work done by a force accelerating an atom. (d)
Atoms or ions in their ground state $|b\rangle$ are accelerated through a single-mode
microwave or optical cavity. }
\end{figure}

Furthermore, it is known that many effects related to the interaction of atoms with the
electromagnetic field acquire new features or are enhanced in the cavity quantum
electrodynamics (QED) setting, when the atoms are injected into a high-Q cavity.

Thus we were motivated to study a simple \textit{gedanken} experiment with a beam of
atoms in their ground state that are accelerated through a single-mode microwave cavity
\cite{prl}; see Fig.~1. This model is sufficiently simple so that we are able not only
to find the probabilities of a photon emission and absorption by ground state atoms,
but also to solve the density matrix equation for the photons in the cavity mode
interacting with a beam of atoms and to analyze its steady state solutions.

We found that the radiation is thermal (in the typical case) and the effective
``Boltzmann factor" can be much larger than the above exponentially small value
\cite{prl}. In particular, for the above example it is given by $\alpha /2\pi \omega$,
which is of order $\sim 10^{-3}$. Hence, it is many orders of magnitude larger than
that for the usual Unruh effect and is potentially observable.

We find that the mechanism of simultaneous excitation of both field and atom in a
cavity is the same as for the Unruh effect in free space. We show that in both cases it
is the nonadiabatic transition due to the counter-rotating term $\hat{a}_k^+
\hat{\sigma}^+$ in the interaction Hamiltonian (\ref{inter}). The reason for an
enhanced excitation in the cavity is the relatively large amplitude for a quantum
transition $|b,0\rangle \rightarrow |a,1\rangle$ between the dressed atomic states due
to the sudden nonadiabatic switching on of the interaction at the cavity boundaries,
whereas for the Unruh effect in free space the emission is exponentially small due to a
slow switching on. In both cases nonadiabatic effects, however small they are, play a
critical role: there is quite a real emission of a photon accompanied by the excitation
of an atom -- not just dressing of the ground state of an atom as a result of
interaction. As was shown in \cite{prl}, when the cavity boundaries are removed, our
expressions yield the usual Unruh result.

Other processes where counter-rotating terms play a crucial role include, e.g.
parametric resonance and the anomalous Doppler effect. The similarity between the Unruh
effect and the anomalous Doppler effect, in which an atom moving faster than the
velocity of light in a medium is also excited to an upper state while simultaneously
emitting a photon, has been emphasized in \cite{ginzburg}. In both cases the energy for the
excitation of an atom and emission of a photon is taken from the work done by the force supporting the
motion of an atom along the given trajectory.
However, for the
Doppler effect there are no time-changing parameters: the excitation occurs due to the
existence of the Cerenkov resonance.

In section II we formulate the model and write down the Hamiltonian. In section III the
transition probabilities for emission are shown to yield a simple physical picture of
the Unruh radiation. Next, the master equation is derived and the steady-state solution
for the photon density matrix is derived and analyzed. In sections IV-VI we analyze the
mechanism of emission and absorption by the accelerated atoms and its relation to the
standard Unruh effect. the emission and absorption probabilities are calculated. The
resulting integrals are evaluated by the method of stationary phase in all physically
interesting asymptotic limits. The possibility of amplification and laser action is
discussed. The interpretation of the results and discussion are presented in section
VII.

\section{The model}

We start from writing  Hamiltonian for the system consisting of a two-level atom
interacting with the electromagnetic field:
\begin{equation}
\hat{H} = \hat{H}_a + \hat{H}_f + \hat{V}.
\end{equation}
Here $\hat{H}_a = \hbar\omega \hat{\sigma}_z$ is the Hamiltonian for an atom with
ground and excited states $|b>$ and $|a>$, respectively, separated by energy difference
$E_a - E_b = \hbar \omega$, $\hat{\sigma}_z = 1/2 (|a><a| - |b><b|)$ the Pauli matrix,
and $\hat{H}_f = \sum_k \hbar \nu_k \hat{a}^{+}_k \hat{a}_k$ is the field Hamiltonian,
where $ \hat{a}^{+}_k$ and $\hat{a}_k$ are photon creation and annihilation operators
and the k-summation is taken over the electromagnetic modes of a cavity or a free
space, depending on the formulation of the problem.

The atom-field interaction Hamiltonian in the atomic rest frame can be written in the
dipole approximation as
\begin{equation} \label{dipole}
\hat{V} = \hbar \sum_k g_k \left(\hat{a}^{+}_k + \hat{a}_k \right) u_k(z(\tau))
\left(\hat{\sigma} + \hat{\sigma}^+ \right).
\end{equation}
Here $u_k({\bf r})$ form a set of orthogonal, normalized functions,  $\hat{\sigma}^+ =
|a><b|$ and $\hat{\sigma} = |b><a|$ are the atomic raising and lowering operators, and
$g_k = {\bf \mu E'_k}/\hbar$ is the atom-field coupling frequency, which depends on the
atomic dipole moment ${\bf \mu}$ and the electric field amplitude ${\bf E^{\prime}_k}$
in the rest frame of an atom, evaluated on the trajectory $z(\tau)$ of an atom as a
function of proper time $\tau$.

In the interaction representation, the master equation for the density operator
$\hat{\rho}$ can be written as
\begin{equation} \label{me}
i\hbar \frac{d\hat{\rho}}{d\tau} = \left[\hat{V}(\tau) \hat{\rho}\right],
\end{equation}
where the interaction Hamiltonian in this representation can be obtained by replacing
$\hat{a}_k \rightarrow \hat{a}_k \exp(-i \nu_k t(\tau))$, $\hat{a}^+_k \rightarrow
\hat{a}^+_k \exp(i \nu_k t(\tau))$, $\hat{\sigma} \rightarrow \hat{\sigma} \exp(-i
\omega \tau)$, and $\hat{\sigma}^+ \rightarrow \hat{\sigma}^+ \exp(i \omega \tau)$,
where $t$ is the time in the inertial laboratory frame. Note that we do not intend to
use rotating wave approximation since the excitation of an atom from the ground state
with simultaneous emission of a photon is described by counter-rotating terms of the
type $\hat{a}^+ \hat{\sigma}^+$.

Our main goal will be to solve master equation Eq.~(\ref{me}) and analyze its solution
in various limits. However, we can get some insight into the physics of an accelerated
atom-field interaction by solving first a simpler problem, namely, calculating photon
emission and absorption probabilities within the standard first-order perturbation
theory.

\section{Transition probabilities, master equation, and the photon density matrix}

\subsection{Probabilities of photon emission and absorption}

Consider an atom entering the cavity at a proper time $\tau_i$, at which moment the
interaction with a cavity mode is assumed to be turned on. If the interaction is weak
enough, the state vector of the system atom+field at any subsequent time $\tau$ can be
found using the first-order perturbation theory:
\begin{equation}
|\psi(\tau)\rangle = |\psi(\tau_i)\rangle - \frac{i}{\hbar} \int_{\tau_i}^{\tau}
\hat{V}(\tau')\, d \tau'|\psi(\tau_i)\rangle.
\end{equation}
The probability of transition $|\psi(\tau_i)\rangle \rightarrow |\psi(\tau)\rangle$ is
therefore given by
\begin{equation}
P = \frac{1}{\hbar^2} \left| \int_{\tau_i}^{\tau} \langle \psi(\tau)| \hat{V}(\tau')
|\psi(\tau_i)\rangle \, d\tau' \right|^2.
\end{equation}

In particular, if an atom was initially in its ground state $|b>$, the probability of
excitation of an atom with simultaneous photon emission into the $k$th mode can be
calculated as
\begin{equation} \label{pem}
P(1_k,a) = \frac{1}{\hbar^2} \left| \int_{\tau_i}^{\tau} g_k(\tau') \langle 1_k|
\hat{a}^+_k |0_k\rangle \langle a| \hat{\sigma}^+ |b\rangle \, d\tau' \right|^2.
\end{equation}
The probability of photon absorption from the $k$th mode by a ground-state atom, when
there is only photon in this mode, is given by
\begin{equation} \label{pabs}
P(0_k,a) = \frac{1}{\hbar^2} \left| \int_{\tau_i}^{\tau} g_k(\tau') \langle 0_k|
\hat{a}_k |1_k\rangle \langle a| \hat{\sigma}^+ |b\rangle \, d\tau' \right|^2.
\end{equation}

Now consider a uniformly accelerated atom moving along the trajectory \cite{Rindler}
\begin{equation} \label{traj}
t(\tau)= t_0 + \frac{1}{\alpha} \rm{sinh}(\alpha \tau), \quad
z(\tau)=\frac{c}{\alpha}\left[ \rm{cosh}(\alpha \tau) - 1 \right],
\end{equation}
where $t_0 = t(\tau = 0)$ is the moment of time in the laboratory (cavity) frame when
the atom starts its acceleration. Expanding the electromagnetic field in terms of
running waves with wave vectors ${\bf k},\; k_z={\bf k}\cdot{\bf v}/v$, the atom-field
interaction Hamiltonian in the atomic frame is given by
\begin{equation} \label{inter}
\hat{V}(\tau)=\sum_k \hbar g_k(\tau)  \left[\hat{a}_k e^{-i \nu t(\tau) +
ik_zz(\tau)}+\rm{h.c.}\right] \left[\hat{\sigma}e^{-i\omega\tau}+\rm{h.c.}\right].
\end{equation}
For the dipole moment ${\bf \mu}$ oriented in x-direction, the corresponding electric
field amplitude in the atomic frame $E'_x$ is related to the x-component of the
electric field amplitude in the lab frame as $E_x^{\prime} = \sqrt{(c-v)/(c+v)}E_x$.
Since $v=c~{\rm{tanh}}(\alpha \tau)$ for a uniformly accelerated particle, we have
$E_x^{\prime}={\rm{e}}^{-\alpha\tau} E_x$ and $g_k(\tau)=g_k{\rm{e}}^{-\alpha \tau}$.

For simplicity, consider the case of a single-mode cavity and co-propagating atom and
field: $k_z=|{\bf k}|=\nu/c$. Substituting Eqs.~(\ref{traj}), (\ref{int}) into
Eqs.~(\ref{pem}), (\ref{pabs}), we obtain \begin{equation} \label{prob} P(0_k,a) = g^2
|I_a(\omega)|^2; \; P(1_k,a) = g^2 |I_e(\omega)|^2,
\end{equation}
where the absorption and emission amplitudes are given, respectively, by
\begin{equation} \label{int} \begin{array}{l} I_a(\omega)=\int_{\tau_i}^{\tau_e}
{\rm{exp}}\left[i\frac{\nu}{\alpha}\left({\rm{e}}^{-\alpha\tau}-1\right)+
i\omega\tau-\alpha\tau\right]d\tau, \\ I_e(\omega) = I_a(-\omega),
\end{array}
\end{equation}
and it is assumed that an atom enters the cavity at $\tau = \tau_i$ and
exits cavity at $\tau = \tau_e$. Hereafter we skip the index $k$  in the coupling
constant $g_k$ for the single-mode cavity. For a counter-propagating wave one needs to
replace $\alpha \rightarrow -\alpha$ in Eq.~(\ref{int}).

We also note that the amplitude of the process of photon emission by an excited atom is
$I_3 = I_a^*$. Therefore, the probability of this process is $P(1_k,b) = P(0_k,a)$.

If we remove the cavity walls to infinity by letting $\tau_i \rightarrow -\infty$ and $\tau_e \rightarrow
\infty$, the integrals in Eq.~(\ref{int}) are reduced to gamma-functions by making the substitution
$x = \frac{\nu}{\alpha} {\rm e}^{-\alpha \tau}$:
\begin{equation} \label{freespace} I_{a,e} = \frac{1}{\nu}{\rm e}^{-i\frac{\nu}{\alpha}} \left(\frac{\alpha}{\nu}\right)^{\mp i\frac{\omega}{\alpha}}
\int_0^{\infty} {\rm e}^{ix} x^{\mp i \frac{\omega}{\alpha}} \, dx =  \frac{i}{\nu}{\rm e}^{-i\frac{\nu}{\alpha}} \left(\frac{\alpha}{\nu}\right)^{\mp i\frac{\omega}{\alpha}}
{\rm{e}}^{\pm \frac{\pi\omega}{2\alpha}} \Gamma\left(1\mp
\frac{i\omega}{\alpha}\right),
\end{equation}
where the upper and lower signs correspond to the absorption and emission amplitudes respectively.
Using the equality  \begin{equation} \Gamma\left(1- \frac{i\omega}{\alpha}\right)
\Gamma\left(1 + \frac{i\omega}{\alpha}\right) = \frac{\frac{\pi
\omega}{\alpha}}{\sinh\frac{\pi\omega}{\alpha}}, \end{equation} we arrive at
\begin{equation} \label{freespace2}
|I_a|^2 = \frac{1}{\nu^2} {\rm e}^{\frac{\pi\omega}{\alpha}}
\frac{\frac{\pi\omega}{\alpha}}{\sinh\left(\frac{\pi\omega}{\alpha}\right)}
\end{equation}
and
\begin{equation} \label{freespace3}
|I_e|^2 = \frac{1}{\nu^2} {\rm e}^{\frac{-\pi\omega}{\alpha}}
\frac{\frac{\pi\omega}{\alpha}}{\sinh\left(\frac{\pi\omega}{\alpha}\right)} =
\frac{1}{\nu^2} \frac{\frac{2\pi\omega}{\alpha}}{{\rm
e}^{\frac{2\pi\omega}{\alpha}}-1}.
\end{equation}
 Note the familiar Planck factor $\left({\rm e}^{\frac{\hbar\omega}{k_B T_u}}-1\right)^{-1}$ in the emission probability
(\ref{freespace3}) with temperature equal to the Unruh temperature $T_u = \frac{\hbar
a}{2 \pi k_B c}$. The ratio of emission to absorption probabilities is also in
agreement with Unruh result in free space:
\begin{equation} \label{unruh} \frac{P(1_k,a)}{P(0_k,a)} = {\rm e}^{-2\pi\omega/\alpha}.
\end{equation}
It is interesting that the simple one-dimensional model of ref. \cite{prl} and the
present discussion contain the basic physics of the Unruh effect even in the free-space
limit. Strictly speaking, in that limit, the interaction of the atom with all
wavevectors ${\bf k}$ of the vacuum field needs to be taken into account. In Sec.~IV
and V we provide a detailed interpretation of this result and also generalize the
treatment to include the modes with different orientations of ${\bf k}$.

\subsection{Master equation for the density matrix}

Consider again uniformly accelerated atoms moving along a trajectory (\ref{traj}) and
interacting with a single-mode field in a cavity. Our goal is to find the solution to
the master equation Eq.~(\ref{me}) within the perturbation theory following the
approach described in \cite{Lamb,micromaser}. In particular, we will find the
 steady state number
of photons in a cavity mode as a result of interaction with a beam of atoms.
\\

{\it Atom in free space.}

First, we derive the results for a single atom in free space within the approach of the
quantum theory of the laser \cite{Lamb,micromaser}. Equation (\ref{me}) can be
rewritten in the form convenient to apply perturbation theory expansion:
\begin{equation} \label{me2}
\frac{d\rho}{d\tau} = -\frac{i}{\hbar} \left[\hat{V},\rho\right] +
\left(-\frac{i}{\hbar}\right)^2 \int_0^{\tau}
\left[\hat{V}(\tau),\left[\hat{V}(\tau'),\rho(\tau')\right]\right] \, d\tau'.
\end{equation}

In Markov approximation, assuming weak interaction with many field modes we decompose
the density matrix as the product of its atomic and field parts as $\rho(\tau) \simeq
\rho_{atom}(\tau) \otimes \rho_{fld}(0)$. Tracing over the field degrees of freedom we
obtain from Eq.~(\ref{me2})
\begin{eqnarray}
\frac{d\rho_{atom}}{d\tau} = -\frac{1}{\hbar^2} \int_0^{\tau} {\rm Tr_{fld}}
\left[\hat{V}(\tau)\hat{V}(\tau') \rho_{atom}(\tau)\rho_{fld}(0)+ \rho_{atom}(\tau)
\rho_{fld}(0) \hat{V}(\tau') \hat{V}(\tau) \right. & & \nonumber \\
\left.  - \hat{V}(\tau)\rho_{atom}(\tau)\rho_{fld}(0) \hat{V}(\tau') -
\hat{V}(\tau')\rho_{atom}(\tau)\rho_{fld}(0) \hat{V}(\tau) \right] \, d\tau'. & &
\label{me3}
\end{eqnarray}

Let us first consider an atom at rest, when $\tau = t$. In the interaction
representation, using Eq.~(\ref{dipole}) and the replacement described after
Eq.~(\ref{me}), we arrive at the following equation for population of state $|a>$:
\begin{eqnarray}
\frac{d\rho_{aa}}{dt} = -\frac{\mu^2}{\hbar^2} \int_0^{t} dt' \left[ \left( \langle
\sum_k E_k^2 \hat{a}_k(t) \hat{a}^+_k(t') \rangle {\rm e}^{i \omega (t-t')} + \left\{t
\leftrightarrow t' \right\} \right) \rho_{aa} \right. & & \nonumber \\
\left. - \left( \langle \sum_k E_k^2 \hat{a}_k^+(t') \hat{a}_k(t) \rangle {\rm e}^{i
\omega (t-t')} + \left\{t \leftrightarrow t' \right\} \right) \rho_{bb} \right], & &
\label{me4}
\end{eqnarray}
where the field operators in angular brackets can be expressed via average number of
photons in the $k$th mode as
\begin{equation} \begin{array}{l}
\langle  \hat{a}_k^+(t') \hat{a}_k(t) \rangle = \bar{n}_k {\rm e}^{-i \nu_k (t-t')},
\\
\langle \hat{a}_k(t) \hat{a}^+_k(t') \rangle = (1+\bar{n}_k) {\rm e}^{-i \nu_k (t-t')}.
\end{array} \end{equation}

 After performing integration, we finally get Eq.~(\ref{me4}) in the form
\begin{equation} \label{me5}
\frac{d\rho_{aa}}{dt} = - {\rm const}\times \sum_k \left[ (1+ \bar{n}_k) \rho_{aa}(t) +
\bar{n}_k \rho_{bb}(t) \right].
\end{equation}
Its steady-state solution when $\bar{n}_k$ is a thermal field is $\rho_{aa}/\rho_{bb} =
{\rm e}^{-\hbar \omega/kT}$.

For an accelerated atom in Minkowski vacuum one can obtain a familiar Unruh result. One
can also obtain a more general result for an atom accelerated through a thermal (not
vacuum) background electromagnetic field with photon distribution $n_k$:
\begin{eqnarray}
\frac{d\rho_{aa}}{dt}& = & -\frac{\mu^2}{\hbar^2} \int_0^{\tau} d\tau'  \sum_k E_k^2
\langle n_k | (\hat{a}_k + \hat{a}_k^+)_\tau (\hat{a}_k + \hat{a}_k^+)_\tau' | n_k
\rangle \left[ {\rm e}^{i \omega (\tau-\tau')} \rho_{aa} -{\rm e}^{-i \omega
(\tau-\tau')} \rho_{bb} \right] + \; {\rm c.c.} \nonumber \\
&=& -\frac{\mu^2}{\hbar^2} \int_0^{\tau} d\tau' \sum_k \left\{ \bar{n}_k {\rm e}^{i
\nu_k (t-t') - i k (z(t) - z(t'))} + (1 + \bar{n}_k) {\rm e}^{- i \nu_k (t-t') + i k
(z(t) - z(t'))} \right\} \nonumber \\ & & \left[{\rm e}^{i \omega (\tau-\tau')}
\rho_{aa} -{\rm e}^{-i \omega (\tau-\tau')} \rho_{bb} \right],  \label{me6}
\end{eqnarray}
where $t = t(\tau), t' = t(\tau')$.

Next, we proceed following the method described in e.g. P. Milonni or J. Audretsch and
R. M\"{u}ller \cite{2-7}. Namely, we assume that the frequency $\nu_k$ has a small
imaginary part, substitute the equations for a uniformly accelerated trajectory
$t(\tau), z(\tau)$ and perform a summation over $k$ which leads to
\begin{eqnarray}
\frac{d\rho_{aa}}{d\tau} =  -\frac{\mu^2}{\hbar^2} \int_0^{\tau} d\tau' \left\{
\frac{\bar{n}_{k_0}}{(\sinh(a(\tau-\tau')/c + i a \epsilon/c))^2} +
\frac{1+\bar{n}_{k_0}}{(\sinh(a(\tau-\tau')/c - i a \epsilon/c))^2} \right\} & &
\nonumber \\
 \left[{\rm e}^{i \omega (\tau-\tau')} \rho_{aa} -{\rm e}^{-i \omega
(\tau-\tau')} \rho_{bb} \right], & & \label{me7}
\end{eqnarray}
where $k_0 = \omega/c$. Then we represent functions $1/(\sinh x)^2$ as infinite series
$\sum_p \frac{1}{(a (\tau-\tau')/c - \pi i p \pm i \epsilon)^2}$, expand time
integration over infinite limits, evaluate the integrals by method of residues and
finally arrive at
\begin{equation} \label{me8}
\frac{d\rho_{aa}}{d\tau} = - \beta \left[ \bar{n}_T \bar{n}_A + (\bar{n}_T + 1)(
\bar{n}_A + 1) \right] \rho_{aa}(\tau) + \beta \left[ \bar{n}_T (\bar{n}_A +1) +
(\bar{n}_T + 1) \bar{n}_A \right] \rho_{bb}(\tau),
\end{equation}
where
\begin{equation} \label{at}
n_A = \frac{1}{{\rm e}^{\frac{2 \pi \omega}{\alpha}} -1}; \; n_T = \frac{1}{{\rm
e}^{\frac{\hbar \omega}{kT}} -1},
\end{equation}
and $\beta$ is a constant which is unimportant for a steady-state distribution of
populations.

Equation (\ref{me8}) allows one to find steady-state atomic populations for a general
case of an atom accelerated through a thermal field background.
\\

{\it Beam of atoms accelerated through a single-mode cavity.}

In this part our goal is to evaluate the steady-state number of photons in a cavity
mode as a result of interaction with a beam of atoms.

As in the quantum theory of the laser \cite{Lamb,micromaser}, the (microscopic) change
in the density matrix of a cavity mode due to any one atom, $\delta \rho^{i}$, is
small. The (macroscopic) change due to $\Delta N$ atoms is then $\Delta \rho = \sum_i
\delta\rho^{i} = \Delta N \delta\rho$. Writing $\Delta N = r \Delta t$, where $r$ is
the atomic injection rate, we have a coarse grained equation of motion: $\Delta \rho
/\Delta t = r \delta \rho$. The change $\delta \rho^i$ due to an atom injected at time
$\tau_i$ in the atomic rest frame is
\begin{eqnarray} \label{master}
\delta \rho^i =- \frac{1}{\hbar^2} \int_{\tau_i}^{\tau_e}
\int_{\tau_i}^{\tau_i+\tau^{\prime}} {\rm{Tr_{\rm atom}}} \times
\\ \nonumber \times \left[\hat{V}(\tau^{\prime}),\left[\hat{V}(\tau^{\prime
\prime}), \rho^{atom}(\tau_i)\otimes \rho(t(\tau_i)) \right]
\right]d\tau^{\prime}d\tau^{\prime \prime},
\end{eqnarray}
where tr$_{\rm atom}$ denotes the trace over atom states. The time $\tau$ is the atomic
proper time, i.e., the time measured by an observer riding along with the atom. For
simplicity, consider again the case of the co-propagating atom and field and the
interaction Hamiltonian given by (\ref{inter}).

   In the case of random injection times, the equation of motion
for the density matrix of the field is
\begin{eqnarray} \label{rho1}
d\rho_{n,n}/dt ~=&& -R_2
\left[(n+1)\rho_{n,n}-n\rho_{n-1,n-1}\right]\\
\nonumber &&-R_1 \left[n\rho_{n,n}-(n+1)\rho_{n+1,n+1}\right],
\end{eqnarray}
where $R_{1,2}$ are defined in the following. If $R_1>R_2$, there is a steady state
solution which is thermal \cite{Lamb}
\begin{subequations}
\begin{equation}
\rho_{n,n}=e^{-\hbar\nu n/k_BT}\left(1-{\rm{e}}^{-\hbar\nu/k_B}\right),
\end{equation}
\begin{equation}
{\bar{n}} =\sum_nn\rho_{nn}=\frac{1}{{\rm{e}}^{\hbar\nu/k_BT}-1}, \;
{\rm{e}}^{-\hbar\nu/k_B}=\frac{R_2}{R_1},
\end{equation}
\end{subequations}
where an effective temperature of the field in the cavity is $T =\hbar \nu/k_B{\rm
{ln}}\left[R_1/R_2\right]$. Thus, spontaneous emission of randomly injected ground
state atoms in the cavity results in thermal statistics of the mode excitation. Note,
that the thermal statistics of the atomic excitation in the standard Unruh effect in
free space is due to spontaneous emission into a vacuum field reservoir with a
continuous spectrum of modes.

Absorption and emission coefficients $R_{1,2}=r|gI_{1,2}|^2$ are determined by the
amplitudes $g I_{1,2}= -\frac{i}{\hbar} \int_{\tau_i}^{\tau_i+T} V_{1,2} d\tau$ of the
matrix elements $V_1=\langle a,0|\hat{V}|b,1\rangle$ and $V_2=\langle
a,1|\hat{V}|b,0\rangle$ of the interaction Hamiltonian (\ref{inter}), respectively, and
their explicit form is given by Eq.~(\ref{int}). Using the results of the previous
section, we get the same result that in the limit $\nu,  \omega \gg \alpha$ the
emission/absorption ratio is $R_2/R_1 \simeq \alpha/(2\pi\omega)$, which is an
enhancement by many orders of magnitude as compared to the exponentially small value
$R_2/R_1 = \exp(-2\pi\omega/\alpha)$.

\section{Emission and absorption of radiation by ground-state atoms}

After the substitution of variables $x = \frac{\nu}{\alpha} {\rm e}^{-\alpha \tau}$,
the absorption and emission amplitudes can be expressed via incomplete gamma-functions:
\begin{equation} \label{gamma}
I_{a,e}(\omega)=\frac{i}{\nu}\left(\frac{\alpha}{\nu}\right)^{\mp
i\frac{\omega}{\alpha}}{\rm{e}}^{\pm \frac{\pi \omega}{2\alpha} - i\frac{\nu}{\alpha}}
\left[\Gamma(\xi,u {\rm e}^{-\alpha (\tau_e-\tau_i)})-\Gamma(\xi, u)\right],
\end{equation}
where $\xi=1\mp i\frac{\omega}{\alpha}$, $u = - i\frac{\nu}{\alpha}$e$^{-\alpha
\tau_i}$, and $\Gamma(\xi,u) = \int_u^{\infty} {\rm e}^{-x}x^{\xi-1}dx$ is the
incomplete gamma function.

In principle, expressions (\ref{gamma}) can be fully analyzed because the properties
and asymptotic behavior of incomplete gamma-functions are well known. Some
representative graphs of the emission and absorption amplitudes as functions of the
field frequency will be shown below in Figs. 2,3,4. However, it is more instructive and
transparent to directly calculate the asymptotic of the integral (\ref{int}) by
applying integration by parts and the method of stationary phase.

In particular, we consider the most realistic case $\nu, \omega \gg \alpha$ and apply
the stationary phase method that can be summarized as
\begin{equation} \label{sphase1}
\int_a^b F(\tau) {\rm e}^{i A f(\tau)} d\tau = B + S,
\end{equation}
where
\begin{equation} \label{sphase2}
B = \left. \frac{F(\tau){\rm e}^{i A f(\tau)}}{i A f'(\tau)} \right|_a^b + \sum_{n =
1}^N \left. \frac{1}{iA^{n+1}} \left( \frac{-1}{f'(\tau)} \frac{d}{d\tau} \right)^n
\frac{F(\tau) {\rm e}^{i A f(\tau)}}{f'(\tau)} \right|_a^b + o(A^{-N})
\end{equation}
is the contribution from integration boundaries obtained by integration by parts,
\begin{equation} \label{sphase3}
S = \sqrt{\frac{2 \pi i}{A f''(\tau_s)}} \left( F(\tau_s) + O(A^{-1})\right) {\rm e}^{i
A f(\tau_s)}
\end{equation}
is the contribution from a stationary point $\tau_s$ such that $f'(\tau_s) = 0$,
$f''(\tau_s) \neq 0$, obtained by expanding $f(\tau)$ in Taylor series around $\tau_s$.
It is assumed that $A \gg 1$. We will also consider separately the case when the
stationary point approaches one of the integration boundaries; see Eq.~(\ref{general})
below.

Suppose for definiteness that $\nu \geq \omega$ and $\tau_i = 0$.  When $\nu - \omega
\geq \sqrt{\alpha \omega}$, the stationary point $\tau_s = \frac{1}{\alpha}
\log\frac{\nu}{\omega}$ of the absorption integral $I_a$ in Eq.~(\ref{int}) is within
the integration limits and far enough from the boundaries. Therefore, $I_a$ can be
evaluated as a sum of the boundary contribution
\begin{equation} \label{b}
I_{a}^{(b)} \simeq \frac{\exp\left[i\frac{\nu}{\alpha}\left({\rm e}^{-\alpha \tau_e} -
1\right) + i \omega \tau_e - \alpha \tau_e \right]}{-i \nu{\rm e}^{-\alpha \tau_e} +
i\omega} + \frac{1}{i(\nu - \omega)}
\end{equation}
and the contribution from the stationary point $\tau_s$:
\begin{equation} \label{s}
I_{a}^{(s)} \simeq \sqrt{\frac{2 \pi}{|\alpha \omega|}} \frac{\omega}{\nu}
\exp\left(i\frac{\omega - \nu}{\alpha} + i \frac{\omega}{\alpha} \log\frac{\nu}{\omega}
+ i \frac{\pi}{4}\right).
\end{equation}

It is clearly seen that the contribution from the stationary point dominates in the
absorption integral $I_a$. The same result can be of course obtained directly from
Eq.~(\ref{gamma}) after moving the integration boundaries to $\pm \infty$ and
considering the resulting expression
\begin{equation} \label{gamma2}
I_{1}(\omega)=\frac{i}{\nu}\left(\frac{\alpha}{\nu}\right)^{-
i\frac{\omega}{\alpha}}{\rm{e}}^{\frac{\pi \omega}{2\alpha} - i\frac{\nu}{\alpha}}
\Gamma\left(1 - i\frac{\omega}{\alpha}\right)
\end{equation}
in the asymptotic limit of a large complex argument of the gamma-function.

The emission integral $I_e$, which originates from the counter-rotating term $\propto
\hat{a}^+ \hat{\sigma}^+$ in the interaction Hamiltonian, does not have a stationary
point within the integration limits. Therefore, its value is solely determined by the
boundary contribution, and $I_e(\omega) \sim I_{a}^{(b)}(-\omega)$. If we further
assume long enough interaction time, $\alpha \tau_e \gg 1$, the second term on the
right-hand side of (\ref{int}) is much greater than the first term, and we obtain
\begin{equation} \label{ratio1}
\frac{P(1_k,a)}{P(0_k,a)} = \frac{\alpha \nu^2}{2\pi\omega (\nu+\omega)^2},
\end{equation}
which is equal to $\frac{\alpha}{2\pi\omega}$ for $\nu \gg \omega$.

Exactly at resonance, $\nu - \omega = \sqrt{\alpha \omega}$, the stationary point
coincides with the lower integration limit $\tau = 0$. In this case one can show that
the main contribution again comes from the stationary point, and the value of the
integral $I_{1c}$ is two times smaller than (\ref{s}). The resulting ratio of
probabilities is equal to
\begin{equation} \label{ratio2}
\frac{P(1_k,a)}{P(0_k,a)} = \frac{\alpha}{2\pi\omega}.
\end{equation}

The above analysis can be readily generalized for an arbitrary value of $\frac{\nu -
\omega}{\sqrt{\alpha \omega}}$. In this case the stationary phase method gives an
additional term in the integrals $I_{1,2}$ that contains the error function erf[z] of
detuning:
\begin{equation} \label{general}
I_{a}^{(s)} \simeq \sqrt{\frac{2 \pi}{|\alpha \omega|}} \frac{\omega}{2\nu}
\exp\left(i\frac{\omega - \nu}{\alpha} + i \frac{\omega}{\alpha} \log\frac{\nu}{\omega}
+ i \frac{\pi}{4}\right) \left[ 1 + {\rm erf}\left(\frac{\omega - \nu}{\sqrt{2 \alpha
\omega}} {\rm e}^{- i\pi/4}\right)\right].
\end{equation}
The function $|I_{a}^{(s)}|^2$ gives the spectral profile of the absorption line.

The ratio (\ref{ratio1}) or (\ref{ratio2}) is surprisingly large; in fact, it is
exponentially larger than the value
$${\rm e}^{-2\pi\omega/\alpha} = {\rm e}^{-\frac{\hbar \omega}{k_B T_u}}$$
one would expect to obtain on the basis of studies of the Unruh effect. Here
\begin{equation} k_B T_u = \frac{\hbar \alpha}{2\pi} \end{equation} is Unruh temperature.
In our case the effective temperature of radiation in the vacuum state of the cavity
mode is determined from
$$ {\rm e}^{-\frac{\hbar \omega}{k_B T}} = \frac{\alpha}{2\pi\omega}, $$
which gives
\begin{equation} \label{T}
k_B T = \frac{\hbar \omega}{\log\frac{2\pi\omega}{\alpha}}.
\end{equation}

The reason for such a large effective temperature is apparently the sudden turn-on of
the interaction of an atom with a cavity mode. If we eliminate the nonadiabatic
switching effect by letting $\tau_i \rightarrow -\infty$ and $\tau_e \rightarrow
\infty$, the integrals in Eq.~(\ref{gamma}) are reduced to
$$ I_{a,e} = \frac{i}{\nu}{\rm e}^{-i\frac{\nu}{\alpha}} \left(\frac{\alpha}{\nu}\right)^{\mp i\frac{\omega}{\alpha}}
{\rm{e}}^{\pm \frac{\pi\omega}{2\alpha}} \Gamma\left(1\mp
\frac{i\omega}{\alpha}\right).
$$
Using the equality  \begin{equation} \Gamma\left(1- \frac{i\omega}{\alpha}\right)
\Gamma\left(1 + \frac{i\omega}{\alpha}\right) = \frac{\frac{\pi
\omega}{\alpha}}{\sinh\frac{\pi\omega}{\alpha}}, \end{equation} we arrive at the
Unruh-type result
\begin{equation} \label{unruh2} \frac{P(1_k,a)}{P(0_k,a)} = {\rm e}^{-2\pi\omega/\alpha}.
\end{equation}

\section{Angular dependence of emission/absorption probabilities}

The above conclusion does not depend on our assumption of interaction with a single
co-propagating cavity mode and can be generalized for the case of an electromagnetic
mode with an arbitrary ${\bf k}$-vector. Similarly to Sec. III, we calculate the
probability $P(1_{\bf k}, a)$ of excitation of an atom with simultaneous photon
emission into the ${\bf k}$th mode assuming that the field was initially in the vacuum
state. Then we calculate The probability $P(0_{\bf k}, a)$ of photon absorption from
the $k$th mode by a ground-state atom, when there is only photon in this mode. The
arguments of the $P$ functions denote the final state of the field and atom. The ratio
of these probabilities is given by
\begin{equation} \label{ratio3}
\frac{P(1_{\bf k}, a)}{P(0_{\bf k}, a)} = \frac{I_{\bf k}(-\omega)}{I_{\bf k}(\omega)},
\end{equation}
where
\begin{equation} \label{intk}
I_{\bf k}(\omega) = \int_{\tau_i}^{\tau_e} \frac{k_z}{k} \exp[i \nu t(\tau) - i k_z
z(\tau) - i\omega \tau - \alpha \tau]\, d\tau,
\end{equation}
where $k = |{\bf k}| = \nu/c$. The probability of emission by an atom into all
electromagnetic modes is proportional to $\int I_{\bf k}\, d^3k$. We will be interested
in evaluating the ratio (\ref{ratio3}). Using equations for the trajectory of a
uniformly accelerated atom, we arrive at
\begin{equation} \label{intk2}
I_{\bf k}(\omega) = \frac{k_z}{k} {\rm e}^{i \nu \tau_i + i \frac{c k_z}{\alpha}}
\int_{\tau_i}^{\tau_e} \exp\left[i \frac{\nu}{\alpha} \left(\sinh\alpha \tau -
\frac{k_z}{k}\cosh\alpha\tau \right) - i\omega \tau - \alpha \tau\right]\, d\tau.
\end{equation}

As in the case of co-propagating mode, the above integral can be calculated exactly in
the infinite limits and can be evaluated approximately by the method of stationary
phase in finite limits.

For the infinite integration limits, $\tau_i, \tau_e \rightarrow \infty$, it was shown
in \cite{ginzburg} that it is convenient to change the integration variable to
$$ \beta = \alpha \tau - \eta, $$
where $\tanh\eta = k_z/k$. Then the integral in (\ref{intk2}) can be written as
\begin{equation} \label{intk3}
\int_{-\infty}^{\infty} {\rm e}^{i \kappa_{\perp} \sinh\beta - \xi \beta -
i\frac{\omega}{\alpha} \eta - \eta} \, d\beta = 2 {\rm e}^{ - i\frac{\omega}{\alpha}
\eta - \eta - \frac{\xi \pi i}{2}} K_{\xi}(\kappa_{\perp}),
\end{equation}
where $\kappa_{\perp} = k_{\perp} c/\alpha$, $\xi = 1 + i \omega/\alpha$, and
$K_{\xi}(\kappa_{\perp})$ is McDonald function. Using the above result in
(\ref{ratio3}), we obtain
\begin{equation} \label{ratio4}
\frac{P(1_{\bf k}, a)}{P(0_{\bf k}, a)} = {\rm e}^{-\frac{2\pi\omega}{\alpha}}
\frac{\left|K_{1-i\omega/\alpha}(\kappa_{\perp})\right|^2}{\left|K_{1+i\omega/\alpha}(\kappa_{\perp})\right|^2},
\end{equation}
which is ``almost'' Unruh factor in the limit $\omega/\alpha \gg 1$, since $K_{-p}(x) =
K_{p}(x)$. The extra factor of 1 in $\xi$ is due to the fact that we are dealing with
photons that have spin 1. This introduced an additional term in the integral as a
result of Lorentz transformation of the field to the atom frame. For the scalar (spin
0) field we would have exactly the thermal Unruh factor.

To evaluate the integrals in finite limits, let us suppose again for definiteness that
$\nu - \omega \gg \sqrt{\alpha \omega}$, $\tau_i = 0$, and $\alpha \tau_e \gg 1$.

The ``counter-rotating'' integral $I_k(-\omega)$ does not have stationary points, and
its value is
$$ |I_k(-\omega)|^2 \sim \frac{1}{(\nu + \omega)^2}. $$
The integral $I_k(\omega)$ is dominated by contribution from the stationary point
$\tau_s$ defined by
\begin{equation}
\cosh\alpha\tau_s - \frac{k_z}{k} \sinh\alpha\tau_s = \frac{\omega}{\nu}.
\end{equation}
It is easy to find that
\begin{equation}
I_k(\omega) \simeq \frac{k_z}{k} \sqrt{\frac{2 \pi}{\alpha \sqrt{\omega^2 - k_{\perp}^2
c^2}}} \; {\rm e}^{i\frac{\sqrt{\omega^2 - k_{\perp}^2c^2}}{\alpha}+ i \frac{c
k_z}{\alpha} - i \omega \tau_s -\alpha \tau_s + i \frac{\pi}{4}}.
\end{equation}
As in the co-propagating mode case, the ratio (\ref{ratio3}) is anomalously large: it
is not exponentially small but linear with respect to $\alpha/\omega$.

\section{Counter-resonant gain and parametric amplification}

Remarkably, not only enhanced spontaneous emission but also laser gain and parametric
gain are possible in cavity QED via counter-resonant emission by ground state atoms
even with random injection times. The gain is reached when $|I_e/I_a|^2 > 1$, or, more
exactly, an excess in $|I_e/I_a|^2$ over 1 should be greater than the normalized cavity
losses. For the gain to occur, the time of flight $T$ should be within a certain range
to ensure that the atom emits into the cavity mode more energy than it takes away,
$R_2>R_1$.

In the case of uniformly accelerated atoms, we find that for a co-propagating
 wave the gain is possible only when the acceleration is large enough: $\alpha >
 \omega$. In the opposite case the ratio $R_2/R_1$ approaches the asymptotic value of
 $\alpha/2\pi \omega$, as was shown in previous sections. Below we plot the ratio
 $R_2/R_1$ for both cases using the incomplete gamma-function representation of
 emission and absorption integrals (\ref{gamma}). Instead of varying the time of flight
 $T$, we plot the gain spectrum as a function of the electromagnetic field frequency
 $\nu$ for the fixed values of $T$ and the atomic frequency $\omega$.

 As is seen from Fig.~2, when $\omega/\alpha >1$, the emission to absorption ratio drops
 down to almost zero due to a large absorption near the resonance frequency $\nu = \omega$ and
 then approaches the asymptotic value $\alpha/2\pi \omega$ in the oscillatory way. When
$\omega/\alpha < 1$, there are strong peaks of a large ratio $R_2/R_1 \gg 1$ at
frequencies corresponding to minima of the absorption probability; see Fig.~3. Note
that the minima of the emission rate are shifted with respect to the minima in the
absorption. At large field frequencies the envelope of the emission to absorption rate
peaks approaches the asymptotic value $2(2\alpha /\pi \omega)^2$.

\begin{figure}[ht]
\centerline{\includegraphics[scale=0.6]{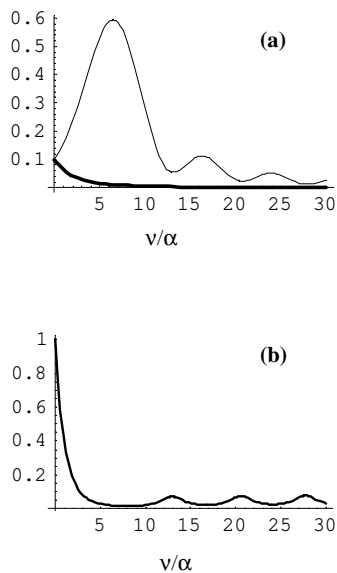}} \caption{\label{Fig02} (a) Absorption
rate $|I_a|^2$ (thin line) and emission rate $|I_e|^2$ (thick line) as functions of the
field-to-acceleration frequency ratio  $\nu/\alpha$  for the atomic frequency $\omega =
3\alpha$  and co-propagating wave. Integrals $I_{a,e}$ are given by Eq.~(11). (b) The
ratio of emission and absorption rates shown in (a). At large $\nu/\alpha$ the curve
reaches the asymptotic value  $\alpha/2\pi \omega$.}
\end{figure}

\begin{figure}[ht]
\centerline{\includegraphics[scale=0.6]{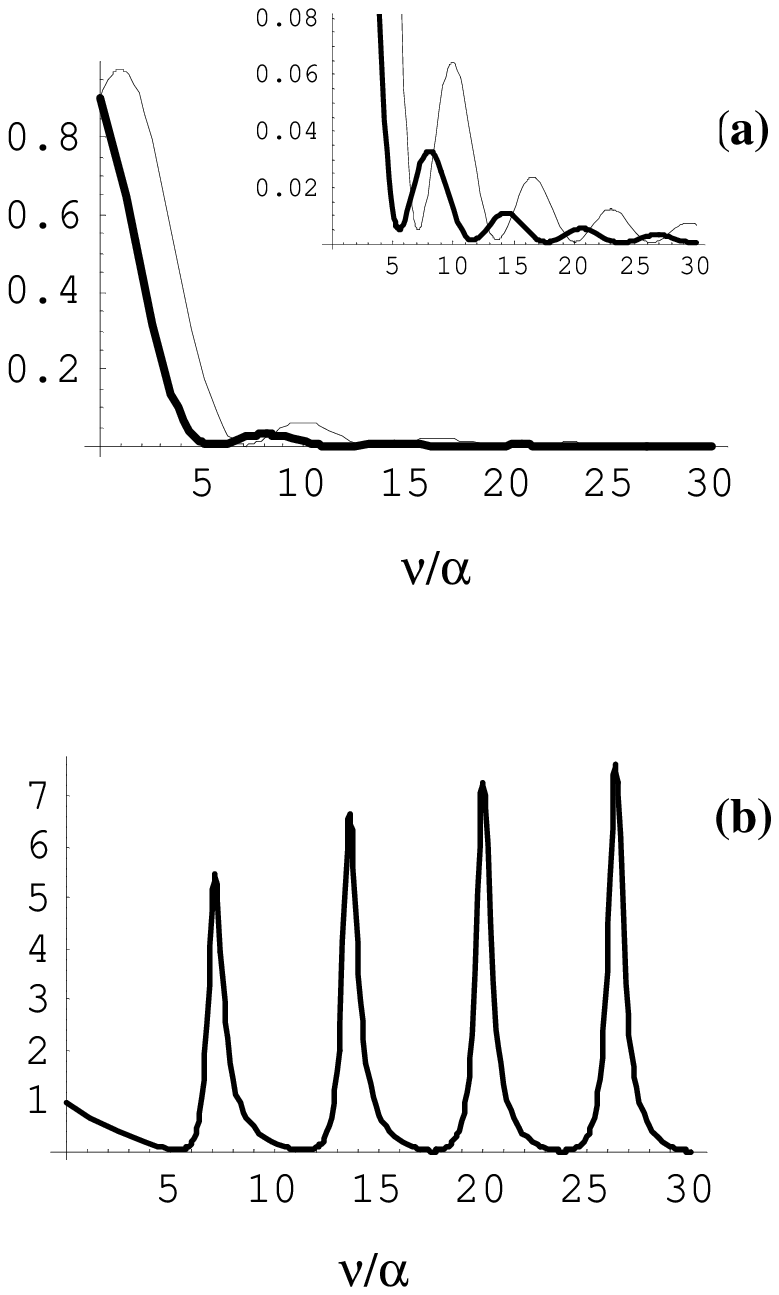}} \caption{\label{Fig03} (a) Absorption
rate $|I_a|^2$ (thin line) and emission rate $|I_e|^2$ (thick line) as functions of the
field-to-acceleration frequency ratio  $\nu/\alpha$  for the atomic frequency $\omega =
3\alpha$  and co-propagating wave. Integrals $I_{a,e}$ are given by Eq.~(11). (b) The
ratio of emission and absorption rates shown in (a). At large $\nu/\alpha$ the curve
reaches the asymptotic value  $\alpha/2\pi \omega$.}
\end{figure}

\begin{figure}[ht]
\centerline{\includegraphics[scale=0.6]{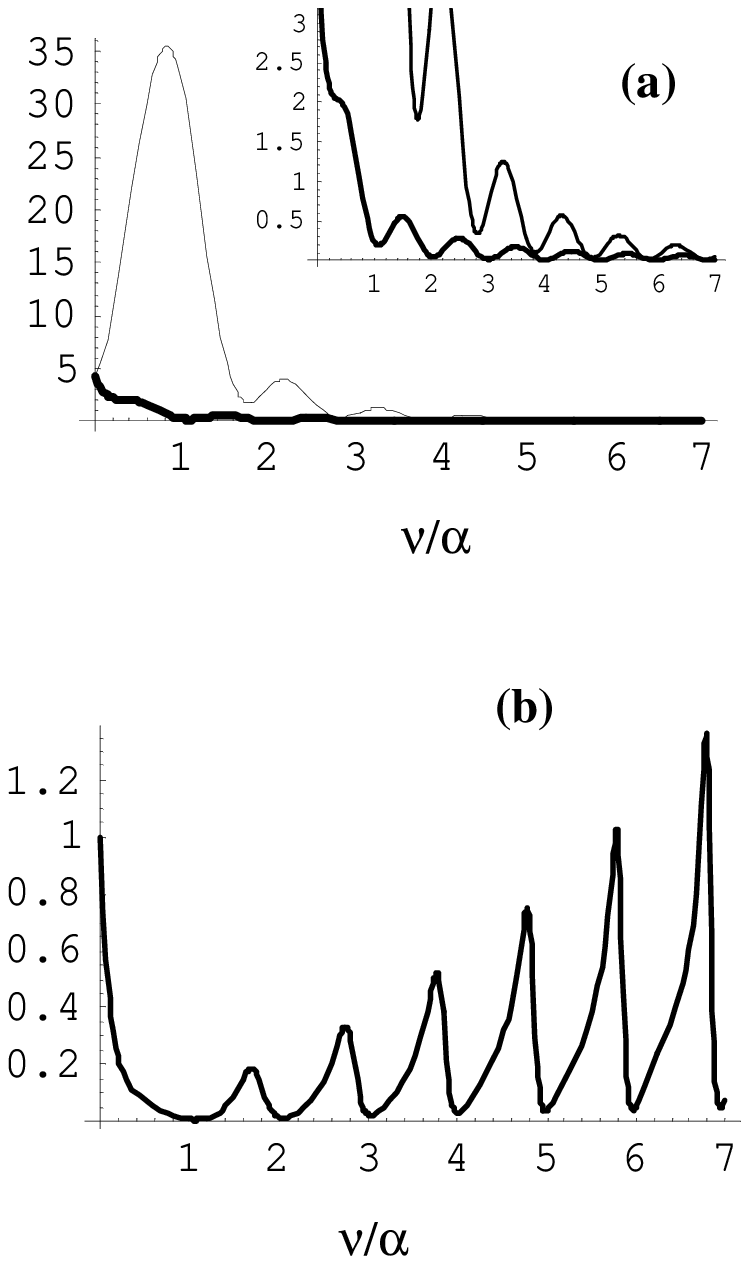}} \caption{\label{Fig04} (a) Absorption
rate $|I_a|^2$ (thin line) and emission rate $|I_e|^2$ (thick line) as functions of the
field-to-acceleration frequency ratio  $\nu/\alpha$  for the atomic frequency $\omega =
3\alpha$  and co-propagating wave. Integrals $I_{a,e}$ are given by Eq.~(11). (b) The
ratio of emission and absorption rates shown in (a). At large $\nu/\alpha$ the curve
reaches the asymptotic value  $\alpha/2\pi \omega$.}
\end{figure}

The counter-propagating mode is more favorable for the amplification due to sharp dips
in the absorption spectrum. As is illustrated in Fig.~4, even in the limit $\omega \gg
\alpha$ the gain spectrum has sharp maxima larger than 1 at the points corresponding to
nearly vanishing absorption. The case $\omega < \alpha$ is qualitatively similar to
that of a co-propagating mode.

Note that the peaks of large gain in Figs.~3,4 are not due to maxima of the emission
integral but due to minima of the absorption probability that are shifted with respect
to the minima of the emission spectrum. Absolute values of both integrals are small.
This is illustrated in the insets to Figs.~3,4 where the emission and absorption
spectra are shown on the same plot.

In the optimal regime for amplification, when $\omega \sim \alpha$, $\nu \gg \alpha$,
and e$^{-\alpha T} \ll 1$ where the time of flight $T \simeq L/c$, one needs to use a
longitudinal cavity mode $\Omega_n = n \pi c/L$ with index $n > 1$. For example, if
$\alpha T \simeq \alpha L/c = 10$, to provide $\nu = \Omega_n = n \pi c/L = 10 \alpha$
one needs $n \simeq 3$.  The multimode regime is possible. It is expected to give the
same qualitative results.

The effects originated from counter-rotating terms are in fact not uncommon. Two
well-known examples are parametric resonance and anomalous Doppler effect
\cite{ginzburg}. In all ``counter-rotating'' processes, an atom can emit a photon and
simultaneously make a transition from ground to excited state. The required energy is
provided by the work done by an external force that sustains the center-of-mass motion
of an atom along a given trajectory. However, an important difference between the
nonadiabatic  processes considered in this paper and the anomalous Doppler effect is
that the latter does not require any time-changing parameters.

It is clear from the above derivation of the emission and absorption probabilities that
the enhancement of the acceleration radiation is related to a strong nonadiabatic
effect at the cavity boundaries. Evidently, this effect should exist for an arbitrary
trajectory of an atom and in particular, for an atom moving with constant velocity. Of
course, the presence of acceleration leads to both  qualitative and quantitative
changes in the excitation rate and emission/absorption probabilities by allowing the
atom to pass through the resonance between the transition frequency of the atom and the
Doppler-shifted frequency of the field.

For a ground-state atom moving through a cavity with a constant velocity and
interacting with a co-propagating wave, it is straightforward to obtain the analytic
expressions for $R_2$ and $R_1$:
\begin{equation} \label{vconst1}
R_1=g^2 \left(\frac{1}{\nu^{\prime}-\omega}\right)^2
\left|1-e^{-i(\nu^{\prime}-\omega)T}\right|^2,
\end{equation}
\begin{equation} \label{vconst2}
R_2=g^2 \left(\frac{1}{\nu^{\prime}+\omega}\right)^2
\left|1-e^{-i(\nu^{\prime}+\omega)T}\right|^2, \nu^{\prime}=\nu \left(\frac{\nu-{\bf
k}\cdot{\bf v}}{\nu+{\bf k}\cdot{\bf v}}\right)^{1/2}.
\end{equation}

Clearly, the factors $1/(\nu' - \omega)^2$ and $1/(\nu' + \omega)^2$ have the same
origin as the nonadiabatic boundary contribution to the emission and absorption
probabilities given by Eq.~(\ref{b}). When we are far from resonance $\nu' = \omega$,
the magnitudes of $R_{1,2}$ in the cases of constant velocity and constant acceleration
are similar and are proportional to the above factors. Thus, when the acceleration
shifts the frequency $\nu'$ further away from the resonance (e.g. when $\nu < \omega$
for the co-propagating wave or when $\nu > \omega$ for the counter-propagating wave),
the emission-to-absorption ratio is increasing. In this case the effect of acceleration
results in the increase of the steady-state number of photons in the cavity as compared
to the constant velocity case. This tendency is of course reversed when the frequency
is shifted towards the resonance by acceleration.  At the same time, in the case of a
constant acceleration we can also have the situation when the atom starts far from
resonance, then passes through resonance in the course of acceleration, and finally
ends up far from the resonance. In this case the ratio $R_2/R_1$ can be quite large and
given by $\alpha/2\pi\omega$, while for an atom moving with a constant velocity and
close to resonance $|\nu' - \omega| \ll \nu',\omega$ the ratio $R_2/R_1$ is very small
due to a strongly enhanced absorption. Thus, depending on the initial conditions,
acceleration can lead to either increase or decrease in the emission-to-absorption
ratio.

The right-hand side of Eqs. (\ref{vconst1}),(\ref{vconst2})  strongly depends also on
the interference factors $e^{-i(\nu^{\prime}\mp \omega)T}$ that are defined by the time
of flight $T$, i.e. the phase an atom accumulates relative to the cavity mode while
passing through the cavity. The ratio $R_2/R_1$ can be even greater than one. To
achieve $R_2/R_1 > 1$, one can tune the time of flight to get the proper interference
factors: $e^{-i(\nu^{\prime}-\omega)T} \rightarrow 1, \;
|e^{-i(\nu^{\prime}+\omega)T}-1| \sim 1$. A similar time of flight tuning is used in
some electronic devices, e.g., in klystrons. The above requirements define a set of the
time-of-flight values, with the maximum gain corresponding to
\begin{equation} \label{tof1} \begin{array}{l}
(\nu - kv + \omega) T = (2n_1 - 1) \pi; \\
(\nu - kv - \omega) T = 2 n_2 \pi,
\end{array}
\end{equation}
where $n_{1,2}$ are integer numbers. For the particular case $n_1 = 0, n_2 = - 1$ one
obtains $ 2 \omega T = \pi$. The monochromaticity of the beam should satisfy the
condition
$$ \frac{\Delta v}{v} \sim \frac{\Delta T}{T} \ll \frac{\pi}{2 \omega T} \sim
\frac{v}{4 c} \frac{\lambda}{L}, $$ where $\lambda = 2\pi c/\omega$, $L$ is a cavity
length, and we assumed $\omega \sim \nu \gg kv$. For $v \sim 1$ km/s and $L \sim
\lambda$ one gets $\Delta v/v \ll 10^{-6}$, which is tough but possible to satisfy.

The counter-propagating mode is more favorable for the gain since the absorption
 can then be anomalously small while the gain remains as large as for the
co-propagating mode.

Similar interference effects, obviously, are present in the case of a constant
acceleration according to Eqs.~(\ref{int}), (\ref{gamma}), (\ref{b}), and (\ref{s}), as
can be seen in Figs.~1-3. They can also lead to the net gain, as we have already
discussed.

In the case of a parametric resonance, consider an atom moving along an oscillating
trajectory $z = z_0 + A\cos\omega_0 t$, $t = \tau$. The photon  absorption and emission
probabilities by a ground-state atom (\ref{pem}) are given by
\begin{equation} \label{param}
R_{1,2}  = g^2 \left| \int_{0}^{\tau_e} {\rm e}^{- i k_z z + i \nu t \mp i \omega t -
\gamma t }\, dt \right|^2, \end{equation} where we introduced a small factor $\gamma$
describing the atomic decay. Using
$$ {\rm e}^{i k A \cos\omega_0 t} = \sum_{p = -\infty}^{\infty} i^p J_p(k_zA) {\rm e}^{i
p \omega_0 t}, $$ the above probabilities can be written as
\begin{equation} \label{param2}
R_{1,2} =  \left|\sum_{p = -\infty}^{\infty} \frac{g J_p(k_zA)}{p\omega_0 \mp \omega +
\nu + i \gamma}\right|^2,
\end{equation}
where $J_p(x)$ is Bessel's function. Evidently, the probabilities are sharply peaked
close to parametric resonance, where $p\omega_0 \pm \omega + \nu_k \simeq 0$. Resonance
for emission corresponds to $\nu + \omega = p \omega_0$, while the absorption resonance
is at $\nu - \omega = p \omega_0$. When $\omega = \nu$,  absorption is always stronger
than emission. Indeed, resonance in absorption exists for $p = 0$, while parametric
resonance in emission requires $p \geq 1$. Therefore, in this case $R_2/R_1 \sim
J^2_p(k_z A)/J^2_0(k_z A) < 1$ for $p \geq 1$. However, when an atom is not at
resonance with the field, one can have parametric resonance in emission but no
resonance in absorption, which results in the parametric gain. The energy is drawn from
the external force causing an atom to follow an oscillating trajectory, and the high
efficiency of this energy transfer is due to a non-stationary, strongly nonadiabatic
character of the atomic center-of-mass motion. In the case of Unruh effect, i.e. a
uniformly accelerated atom in free space, it is also nonadiabaticity that drives
simultaneous excitation of the atom and the field. However, the efficiency is much
lower due to much slower change in the atomic velocity. For an atom entering the
cavity, a sudden nonadiabatic switch-on of the interaction causes a stronger
excitation.

\section{Nonadiabatic nature of acceleration radiation}

   The above calculations clearly show that the mechanism of simultaneous excitation of
both field and atom is the same as for the Unruh effect in free space, namely
nonadiabatic transition due to the counter-rotating term $\hat{a}_k^+ \hat{\sigma}^+$
in the interaction Hamiltonian (\ref{inter}), i.e. $V_2$. The reason for an enhanced
excitation in the cavity is the relatively large amplitude for a quantum transition
$|b,0\rangle \rightarrow |a,1\rangle$ due to the sudden nonadiabatic switching on of
the interaction, whereas for the Unruh effect in free space the emission is
exponentially small due to a slow switching on. However, in both cases there is quite a
real emission of a photon accompanied by the excitation of an atom -- not just dressing
of the ground state of an atom as a result of interaction.

We will now illustrate the above statement by explicit derivation of both the Unruh
factor and the enhanced excitation factor as a probability of the nonadiabatic
transition from the dressed ground state to the dressed excited state. Consider first
our case of a sudden turn on of the interaction in cavity QED.  As a result of the
interaction, the initial state $|b,0\rangle$ is no longer an eigenstate of the
Hamiltonian. Now, a linear superposition of the excited states of the atom and field
makes up the dressed \cite{siegert} ground state of the interacting system $ \psi_0 =
|b,0\rangle - \frac{g(\tau)}{\nu^{\prime} + \omega} |a,1\rangle$ as well as the dressed
excited state $ \psi_1 = |a,1\rangle + \frac{g(\tau)}{\nu^{\prime} + \omega}
|b,0\rangle$.

In particular, the amplitude of the bare excited state $|a,1\rangle$ in $\psi_0$ is of
the order of $C \sim \mu E^{\prime}/\hbar(\omega+\nu^{\prime})$. It is easy to
calculate that the latter corresponds to the atomic excitation probability
$\rho_{aa}^{atom} = |C|^2 \sim |\mu E^{\prime}/\hbar(\omega+\nu)|^2 \sim |gI_e|^2$,
where the emission integral $I_e$ is defined above. This result can be also obtained
directly from the density matrix equation for the atom, via the atomic counterpart to
Eq. (\ref{master}) with a trace over the photon states instead of the tr$_{\rm atom}$.
This probability has the same origin and value as the well-known Bloch-Siegert shift of
a two-level atomic transition \cite{siegert}, $\Delta\omega/\omega = (\mu
E^{\prime}/\hbar(\omega+\nu))^2$, due to counter-rotating terms in the interaction
Hamiltonian.

The counter-rotating term in Eq.~(\ref{b}) represents the contribution from boundaries
to the nonadiabatic transition amplitudes. In the absence of the boundary
contributions, the emission integral $I_e(\omega) = I_a(-\omega)$ in Eq.~(\ref{b})
becomes exponentially small $\sim \exp(-\pi\omega/\alpha)$ for the small parameter
$\alpha/2\pi\omega \ll 1$ since there are no stationary phase points in the integration
interval. The absorption integral $I_a$ does have a point of stationary phase when the
atomic frequency $\omega$ is brought into resonance with the field due to the
time-dependent Doppler shift of the mode frequency \cite{doppler}
$\nu^{\prime}(\tau)=\nu\exp(-\alpha\tau)$. This fact explains why the related
exponential factor effectively disappears from the absorption integral (\ref{s}) when
$\alpha \ll 2 \pi \omega$. As a result, if there are no edge effects, we obtain the
same excitation factor $R_2/R_1=\exp(-2\pi\omega/\alpha)$ as in the Unruh effect (in
free space). This means that in order to observe the standard Unruh result one has to
extend the mode profile $g(z)$ near the boundaries, i.e., eliminate nonadiabatic
boundary contributions.

Similary to what we did for the sudden turn-on case,  let us now demonstrate the
nonadiabatic nature of the Unruh effect by the following explicit derivation of the
Unruh factor as a probability of the nonadiabatic transition $\psi_0 \rightarrow
\psi_1$ from the dressed ground state. The Shroedinger equation $i\hbar d\psi/d\tau = H
\psi$ in the two-level case $\psi = c_0\psi_0 + c_1 \psi_1$ yields $dc_1/d\tau + (i
E_1/\hbar + \langle \dot{\psi_1}|\psi_1\rangle) c_1 = - c_{0}\langle
\dot{\psi}_{0}|\psi_1\rangle$. The difference between the eigenenergies is, to the
first order, $E_1 - E_0 = \hbar (\omega + \nu^{\prime})$. For small nonadiabatic
coupling $ -\langle \dot{\psi_0}|\psi_1\rangle =
\frac{d}{d\tau}\left(\frac{g(\tau)}{\omega + \nu^{\prime}}\right) \ll \omega +
\nu^{\prime}$, the perturbation solution is $|c_1|^2 = |\int_{\tau_i}^{\tau} \exp[i
\int_{\tau_i}^{\tau^{\prime}} (\nu^{\prime} + \omega) d\tau'']
\frac{d}{d\tau^{\prime}}\left(\frac{g(\tau^{\prime})}{\omega + \nu^{\prime}}\right)\,
d\tau'|^2$. If we now make the assumption of an adiabatic switching (on and off) of the
interaction $g(\tau)$ as in standard Unruh effect treatments, then after integration by
parts the latter integral is reduced to the integral $I_e(\omega) = I_a(-\omega)$ in
Eqs. (6) but in the infinite limits, i.e. without edge effects. This yields the
standard Unruh factor $|c_1|^2 \propto \exp(-2\pi\omega/\alpha)$. This derivation
clearly shows the dramatic effect of boundary contributions leading to a large
amplitude $\sim g(\tau)/(\omega + \nu')$ of the atomic excited state $|a\rangle$. Only
if we eliminate the edge effects by adiabatic switching of the interaction do we
retrieve the exponentially small excitation factor.

   Note that in the cavity the excitation
factor $\exp(-\hbar\nu/k_BT)\equiv R_2/R_1=\alpha/2\pi\omega$ is determined by the
first power of the same nonadiabaticity parameter $\alpha/2\pi\omega$. The reason for
this effect is the existence of a true resonance, i.e., a stationary-phase point, in
the absorption coefficient. As mentioned earlier, this yields a resonance between the
atomic transition frequency and the Doppler-shifted frequency of the field seen by the
atom, $\omega + \frac{d}{d\tau}(\frac{\nu}{\alpha}{\rm {e}}^{-\alpha\tau}) \simeq 0$,
and is responsible for the aforementioned effect.

\section{Conclusions}

 Our simple model clearly demonstrates that the ground state atoms accelerated
through a vacuum-state cavity radiate real photons. For relatively small acceleration
$a < 2\pi\omega c$, the excitation Boltzman factor ${\rm{exp}}(-\hbar\nu/k_BT)\sim
\alpha / 2 \pi \omega$ is much larger than the standard Unruh factor
$\exp(-2\pi\omega/\alpha)$. The physical origin of the
 field energy in the cavity and of the real internal energy in
the atom is, of course, the work done by an external force driving the center-of-mass
motion of the atom against the radiation reaction force. Both the present effect (in a
cavity) and standard Unruh effect (in free space) originate from the transition of the
ground state atom to the excited state with simultaneous emission of photon due to the
counter-rotating term $\hat{a}_k^+\hat{\sigma}^+$ in the time-dependent Hamiltonian
(9). Thus, these effects have essentially the same counter-resonant, nonadiabatic
mechanism. We emphasize that there is emission of real photons in both cases; however
the emission probability is exponentially small for the standard Unruh condition of the
absence of boundaries and slow turn-on of the interaction -- simply because the
nonadiabatic effect is very small in the latter case. The enhanced rate of emission
into the cavity mode comes from the enhanced nonadiabatic transition at the cavity
boundaries; the standard Unruh excitation comes from the nonadiabatic transition in
free space due to the time dependence of the Doppler-shifted field frequency
$\nu^{\prime}=\nu {\rm e}^{-\alpha\tau}$, as seen by the atom in the course of
acceleration.

The authors gratefully acknowledge the support from DARPA-QuIST, ONR, and the Welch
Foundation. We would also like to thank R. Allen, H. Brandt, I. Cirac, J. Dowling,  R.
Indik, P. Meystre, W. Schleich, L. Susskind, and W. Unruh for helpful discussions.

\newpage

\begin{center}
\begin{large}

Figure Captions

\end{large}
\end{center}

Fig. 1. (a) Atoms or ions in their ground state $|b\rangle$ are accelerated through a
single-mode microwave or optical cavity. (b) Resonant absorption or emission: an atom
is excited (deexcited) as it simultaneously absorbs (emits) a photon. (c)
Counter-resonant absorption or emission processes that are usually neglected in the
``rotating wave approximation'': an atom is excited (deexcited) as it simultaneously
emits (absorbs) a photon. (d) the energy for counter-resonant emission is drawn from
work done by a force accelerating an atom. \\

Fig. 2. (a) Absorption rate $|I_a|^2$ (thin line) and emission rate $|I_e|^2$ (thick
line) as functions of the field-to-acceleration frequency ratio  $\nu/\alpha$  for the
atomic frequency $\omega = 3\alpha$  and co-propagating wave. Integrals $I_{a,e}$ are
given by Eq.~(11). (b) The ratio of emission and absorption rates shown in (a). At
large $\nu/\alpha$ the curve reaches the asymptotic value  $\alpha/2\pi \omega$. \\

Fig. 3. (a) Absorption rate $|I_a|^2$ (thin line) and emission rate $|I_e|^2$ (thick
line) as functions of the field-to-acceleration frequency ratio  $\nu/\alpha$  for the
atomic frequency $\omega = \alpha/3$  and co-propagating wave. Integrals $I_{a,e}$ are
given by Eq.~(11). The inset shows the tails in more detail. (b) The ratio of emission
and absorption rates shown in (a). At
large $\nu/\alpha$ the curve reaches the asymptotic value  $(2\alpha/\pi \omega)^2/2 = 8$. \\

Fig. 4. (a) Absorption rate $|I_a|^2$ (thin line) and emission rate $|I_e|^2$ (thick
line) as functions of the field-to-acceleration frequency ratio  $\nu/\alpha$  for the
atomic frequency $\omega = 3\alpha$  and counter-propagating wave. Integrals $I_{a,e}$
are given by Eq.~(11). The inset shows the tails in more detail.
(b) The ratio of emission and absorption rates shown in (a).  \\


\begin{thebibliography}{xx}
\bibitem{9}
The frequency $\omega$ is lower bounded (for a possible experiment) by cryogenic
technology and the requirement that the effect should not be obscured by ``hot" walls.
\bibitem{doppler}
The Doppler-shifted frequency of the field, as it is seen by the atom is $\nu^{\prime}
=d(\nu t(\tau)-k_zz(\tau))/d\tau = \nu {\rm{e}}^{-\alpha \tau}$. For the case of a
counter-propagating wave one has to change the sign of $\alpha$ in the
${\rm{e}}^{-\alpha\tau}$ and related Eq. (11).
\end{thebibliography}
\end{document}